\documentclass[aps,prl,showpacs,groupedaddress]{revtex4}
\usepackage{amsmath,amsfonts,amssymb,amscd,amsthm,amsbsy,upref}
\numberwithin{equation}{section}

\def\cee{{\mathbb C}}
\def\real{{\mathbb R}}
\def\zed{{\mathbb Z}}
\def\Circ#1{\raise.3ex\hbox{$\mathop{#1}\limits^\circ$}}

\def\op{\operatorname{op}}
\def\Re{\operatorname{Re}}
\def\cH{{\mathcal H}}
\def\L{{\mathcal L}}
\def\ds{\displaystyle}
\def\simto{\ \buildrel \sim\over\to\ }
\begin{document}
\title[Asymptotic Theory for Quantum Bose Systems]
{Asymptotic Theory for Quantum Bose Systems\\
with Many Degrees of Freedom}
\author{Misha Vishik}
\affiliation{Department of Mathematics\\ The University of Texas at Austin\\
Austin, TX 78712-1082, U.S.A.}
\email[]{vishik@math.utexas.edu}
\author{Gennady Berman}
\affiliation{Los Alamos National Laboratory, MS B213\\ 
Los Alamos, NM 87545, U.S.A.}
\email[]{gpb@lanl.gov} 
\date{\today}
\begin{abstract}
We construct asymptotic expansions of Laplace type for the time-dependent
quantum averages for Bose systems with many degrees of freedom,
initially populated in coherent states. These solutions are localized
in phase space, and they are different from the usual oscillating 
asymptotics for the quasi-classical wave functions. These expansions are
valid on any fixed time interval, and caustics do not contribute to the 
asymptotics.
\end{abstract}

\maketitle

In the present paper we develop an asymptotic theory for the partial 
differential equations describing the evolution of averages in coherent states
as $\hbar \to 0$, for Bose sistems with many degrees of freedom.
The crucial observation is that Laplace asymptotics, describing the 
quantities exponentially localized in the parameter of a coherent state and 
not the standard oscillating WKB asymptotics play the central role in the 
theory. Remarkably
the phase and the amplitude in Laplace asymptotic expansion can be constructed 
using the classical Hamilton-Jacobi theory
without a usual difficulty associated with caustics. In a sense the expansion
we describe below corresponds
to a "purely imaginary" WKB with a nonnegative phase. 
We use coherent states as initial wave functions of the evolution.
Coherent states are the closest states to the classical systems because they 
realize the minimum uncertainty relation. 
They were introduced in R.J.~Glauber \cite{gla}. 
Different approaches to the dynamical quasi-classical theory were reviewed 
in \cite{gut}, \cite{reich}. 

Our theory is based on  the algebra of non-commutative operators which was 
discussed in detail in \cite{BS}, \cite{AW}, \cite{CN}, \cite{ST}, \cite{BBH}. 
Mathematical foundations of the theory are thoroughly developed in 
the works of V.P.~Maslov \cite{Mas1}, \cite{Mas2}, 
L.~H\"ormander \cite{Hor} and many other authors. 
The theory we develop below is formulated in a general form and can be
applied to rather general Bose systems with many degrees of freedom.
In particular, it can be used for computing dynamical quantum effects
in complicated Bose-Einstein condensate systems.
\addtocounter{section}{1}
\setcounter{equation}{0}
\subsection*{1.}
We consider a time-independent Hamiltonian 
\begin{equation}\label{eq:1.1}
H(a_1^\dagger  ,\ldots,a_N^\dagger , a_1,\ldots,a_N) = \sum_{\ell,s} H_{\ell s} a^{\dagger \ell}
a^s\ ,
\end{equation}
where  
\begin{equation*}
\ell = (\ell_1,\ldots,\ell_N) \in \zed_+^N\ ,\quad 
s= (s_1,\ldots,s_N) \in \zed_+^N\ ;
\end{equation*} 
$\zed_+$ stands for the set of nonnegative integers; $a^{\dagger \ell} = a_1^{\dagger \ell_1}
\cdots a_N^{\dagger \ell_N}$; $a^s = a_1^{s_1} \cdots a_N^{s_N}$. 
Here as usual 
\begin{equation*}
a_k^\dagger  = \frac1{\sqrt 2} \left( x_k - \hbar \frac{\partial}{\partial x_k}\right)
\ ,\quad 
a_k = \frac1{\sqrt 2} \left( x_k + \hbar \frac{\partial}{\partial x_k}\right)
\end{equation*}
are the creation operator and the annihilation operator respectively; 
$[a_k^\dagger ,a_\ell] = - \delta_{k\ell} \hbar$ for $k,\ell = 1,\ldots,N$. 

We assume $H$ is formally self-adjoint, i.e., 
\begin{equation}\label{eq:1.2} 
H_{\ell s} = H_{s\ell}^*\ .
\end{equation}
The Heisenberg equation for an arbitrary operator valued function $F(t)$ is 
\begin{equation}\label{eq:1.3} 
\dot F = \frac{i}{\hbar} [H,F]\ .
\end{equation}
We introduce Poisson vectors $\Phi_\alpha$ and coherent states 
$|\alpha \rangle$ as follows: 
for any $\alpha = (\alpha_1,\ldots,\alpha_N) \in \cee^N$
\begin{align}
\Phi_\alpha & =  (\pi \hbar)^{-N/4} \exp \left\{ - \frac1{2\hbar} 
\left(x^2 - 2\sqrt2\ x\cdot \alpha + \alpha^2\right)\right\}\ ,
\label{eq:1.4}\\
|\alpha\rangle & =  \exp \left( - \frac{|\alpha|^2}{2\hbar}\right)\Phi_\alpha\ .
\label{eq:1.5}
\end{align}
Then, $a\Phi_\alpha  = \alpha \Phi_\alpha$, i.e., 
\begin{equation}\label{eq:1.6} 
\begin{array}{rcll}
a_k \Phi_\alpha & =& \alpha_k \Phi_\alpha\ ,&\qquad k= 1,\ldots,N\ .
\quad \text{Also}\\
\noalign{\vskip6pt}
\displaystyle \hbar \frac{\partial}{\partial\alpha_k} \Phi_\alpha 
& = & a_k^\dagger  \Phi_\alpha\ ,&\qquad k=1,\ldots,N\ .
\end{array}
\end{equation}
Unlike $\Phi_\alpha$, the vectors $|\alpha\rangle$ are normalized: 
$\langle \alpha\mid\alpha\rangle =1$. 
Notice that $\Phi_\alpha$ is holomorphic with respect to $\alpha\in\cee^N$.
Our goal is to study the expectation value $\langle\alpha |F(t)|\alpha\rangle$.
To obtain the partial differential equation for this quantity we argue as 
follows.

Let $F(t) = \sum_{m,q\in\zed_+^N} F_{mq}(t) a^{\dagger m} a^q$. 
Using \eqref{eq:1.1}, \eqref{eq:1.3}, \eqref{eq:1.6}, 
\begin{align}
&\frac{d}{dt} \langle\Phi_\alpha| F\Phi_\alpha\rangle \label{eq:1.7}\\
& = \frac{i}{\hbar} \langle \Phi_\alpha| (HF - FH)\Phi_\alpha\rangle\notag\\
& = \frac{i}{\hbar} \sum_{\ell,s,m,q} 
H_{\ell s} F_{mq} (t) \left\{ \langle\Phi_\alpha |a^{\dagger \ell} 
a^s a^{\dagger m} a^q \Phi_\alpha\rangle - \langle\Phi_\alpha|a^{\dagger m} a^q a^{\dagger \ell} a^s 
\Phi_\alpha\rangle\right\}\notag\\
& = \frac{i}{\hbar} \sum_{\ell,s,m,q} H_{\ell s} F_{mq} 
\left\{ \alpha^{*\ell} \alpha^q \langle a^{\dagger s} \Phi_\alpha| a^{\dagger m} \Phi_\alpha
\rangle 
- \alpha^{*m} \alpha^s \langle a^{\dagger q} \Phi_\alpha| a^{\dagger \ell} \Phi_\alpha\rangle
\right\}
\notag\\
& = \frac{i}{\hbar} \sum_{\ell,s,m,q} H_{\ell s} F_{mq} 
\left\{ \alpha^{*\ell} \alpha^q \left(\hbar \frac{\partial}{\partial \alpha^*}
\right)^s \left(\hbar\frac{\partial}{\partial \alpha}\right)^m 
 - \alpha^{*m} \alpha^s \left( \hbar \frac{\partial}{\partial \alpha^*}
\right)^q \left( \hbar\frac{\partial}{\partial\alpha}\right)^\ell\right\}
\exp \frac{|\alpha|^2}{\hbar} \notag\\
& = \frac{i}{\hbar} \sum_{\ell,s,m,q} H_{\ell s} F_{mq} 
\left\{ \alpha^{*\ell} \left(\hbar \frac{\partial}{\partial \alpha^*}\right)^s 
\alpha^{*m} \alpha^q - \alpha^s \left(\hbar\frac{\partial}{\partial\alpha}
\right)^\ell \alpha^{*m} \alpha^q\right\} 
\exp \frac{|\alpha|^2}{\hbar}\notag\\
& = \frac{i}{\hbar} \sum_{\ell,s} H_{\ell s} 
\left\{ \alpha^{*\ell} \left(\hbar\frac{\partial}{\partial \alpha^*}\right)^s
- \alpha^s \left(\hbar\frac{\partial}{\partial \alpha}\right)^\ell\right\} 
\exp \frac{|\alpha|^2}{\hbar} 
\langle \alpha |F(t)|\alpha\rangle\ .\notag
\end{align}

In a slightly more general setting, let 
\begin{equation}\label{eq:1.8} 
H = \op^{\text{Wick}} \cH (z^*,z)
\end{equation}
For $H$ as in \eqref{eq:1.1} this means 
\begin{equation*}
\cH (z^*,z) = \sum_{\ell,s} H_{\ell s} z^{*\ell} z^s
\end{equation*}
Then from \eqref{eq:1.5}, \eqref{eq:1.7} we get a closed equation for 
the evolution of $\langle \alpha |F|\alpha\rangle$:
\begin{equation}\label{eq:1.9} 
\begin{split}
&\frac{d}{dt} \langle \alpha |F|\alpha\rangle =\\
&\qquad = \frac{i}{\hbar} \op \left[ \left( \cH \Big(\alpha^* ,\frac12 
\hbar \xi +\alpha \Big) - \cH \Big( \frac12\hbar \xi^*+\alpha^* ,\alpha
\Big)\right)\right]
\langle \alpha |F|\alpha\rangle
\end{split}
\end{equation}

Here, as in the theory of pseudodifferential operators, for any appropriate 
symbol $b(\alpha,\xi)$ 
\begin{equation*}
\op [b(\alpha,\xi)] f= (2\pi)^{-2N} \int_{\cee^N} \int_{\cee^N} 
b(\alpha,\xi) e^{i\Re \xi \cdot (\alpha-\gamma)^*} f(\gamma) \, 
d^{2N} \gamma\,d^{2N} \xi 
\end{equation*}
For the purpose of the present paper it will be sufficient to treat 
the case when $\cH$ is a polynomial in $z^*,z$. 
We will address the mathematical problems related to a more general $\cH$ 
in future publications.

In case $\cH (z^*,z)$ is a polynomial, the equation \eqref{eq:1.9} has 
the following explicit form: 
\begin{equation}\label{eq:1.10} 
\begin{split}
&\frac{d}{dt} \langle \alpha |F|\alpha\rangle = \\
&\qquad = \frac{i}{\hbar} \sum_{r\in\zed_+^N} \frac1{r!} 
\left( \Big(\frac{\partial}{\partial\alpha}\Big)^r \cH (\alpha^*,\alpha) 
\Big(\hbar \frac{\partial}{\partial\alpha^*}\Big)^r 
- \Big(\frac{\partial}{\partial\alpha^*}\Big)^r \cH  
(\alpha^*,\alpha) \Big( \hbar \frac{\partial}{\partial\alpha}\Big)^r \right) 
\langle\alpha|F|\alpha\rangle
\end{split}
\end{equation}
Here, for $r = (r_1,\ldots,r_N)\in\zed_+^N$, $r!= r_1! \cdots r_N!$.

\addtocounter{section}{1}
\setcounter{equation}{0}
\subsection*{2.}	
The goal of the paper is to present an asymptotic solution to 
\eqref{eq:1.9}, \eqref{eq:1.10} as $\hbar \to 0$. 
We first treat the case of the initial condition of the following nature:
\begin{equation}\label{eq:2.1}
\langle\alpha |F(0)|\alpha\rangle 
= \alpha^{*m} \alpha^q \exp \left( - \frac{|\alpha-\alpha_0|^2}{\hbar}
\right)\ ,\qquad \alpha_0 \in \cee^N\ .
\end{equation}
This corresponds, i.e., to the choice 
\begin{equation}\label{eq:2.2}
F(0) = a^{\dagger m} |\alpha_0\rangle \langle \alpha_0 |a^q
\end{equation}
Indeed, for $\alpha \in \cee^N$, 
\begin{equation*}
\begin{split} 
\langle \alpha|a^{\dagger m}|\alpha_0\rangle 
\langle \alpha_0 |a^q|\alpha\rangle 
& = \alpha^{*m}\alpha^q |\langle\alpha_0,\alpha\rangle|^2 \\
& = \alpha^{*m} \alpha^q \left| \exp \left( \frac{\alpha_0^*\alpha}{\hbar}
- \frac{|\alpha_0|^2}{2\hbar} - \frac{|\alpha|^2}{2\hbar}\right)\right|^2\\
& = \alpha^{*m} \alpha^q \exp \left( - \frac{|\alpha-\alpha_0|^2}{\hbar}
\right)\ .
\end{split}
\end{equation*}
To solve asymptotically the problem \eqref{eq:1.9}, \eqref{eq:1.10} with the 
initial condition as in \eqref{eq:2.1} we propose the following procedure.
Let $f(\alpha^*,\alpha,t) = \langle \alpha |F(t)|\alpha\rangle$. 

Assume formally 
\begin{equation}\label{eq:2.3} 
f(\alpha^*,\alpha,t) = e^{\frac{S(\alpha^*,\alpha,t)}{\hbar}} 
\sum_{j=0}^\infty b_j (\alpha^*,\alpha,t) \hbar^j\ ,
\end{equation}
where the phase $S(\alpha^*,\alpha,t)$ and the coefficients 
$b_j$, $j=0,1,2,\ldots$ are to be determined. 

The initial conditions are 
\begin{align}
S(\alpha^*,\alpha,0) & = - |\alpha -\alpha_0|^2\ ;\label{eq:2.4}\\
b_0(\alpha^*,\alpha,0) & = \alpha^{*m} \alpha^q\ ;\label{eq:2.5}\\
b_j(\alpha^*,\alpha,0) & = 0\ \text{ for }\ j\ge 1\ .\label{eq:2.6}
\end{align}
With these choices \eqref{eq:2.1} will follow immediately. 
We substitute \eqref{eq:2.3} into \eqref{eq:1.10} and obtain the 
Hamilton-Jacobi equation for $S$ and a system of transport equations for 
$b_j$, $j\ge 0$. 

For the phase $S$ we have from Taylor's formula: 
\begin{equation}\label{eq:2.7} 
\dot S = i \left\{ \cH \left( 
\alpha^*,\alpha +\frac{\partial S}{\partial\alpha^*}
\right) 
- \cH 
\left( 
\alpha^* +\frac{\partial S}{\partial \alpha},\alpha
\right)\right\}\ .
\end{equation}
It is a rather remarkable fact that the Hamilton-Jacobi equation 
\eqref{eq:2.7} is {\em real\/} (see \eqref{eq:1.2}) and therefore 
the classical Hamilton-Jacobi theory applies. 
The expansion \eqref{eq:2.3} describes Laplace asymptotics unlike WKB 
asymptotics valid for the Schr\"odinger equation (see, e.g., 
\cite{BBH}) describing the evolution of a wave function.

We introduce the momenta $p, p^*$ and the effective Hamiltonian 
associated with \eqref{eq:2.7} as follows: 
\begin{equation}\label{eq:2.8} 
W(\alpha^*,\alpha,p^*,p) 
= - i \left\{ \cH (\alpha^*,\alpha  +p^*) 
- \cH (\alpha^* +p,\alpha)\right\}\ .
\end{equation}

The Hamiltonian dynamics in these variables is 
\begin{equation}\label{eq:2.9}
\left\{
\begin{split}
\dot{\alpha}_k & = i\frac{\partial\cH}{\partial \alpha_k^*} 
(\alpha^* + p,\alpha)\\
\dot{p}_k & = i\left\{ \frac{\partial\cH}{\partial\alpha_k} 
(\alpha^*,\alpha +p^*) - \frac{\partial\cH}{\partial \alpha_k} 
(\alpha^*+p,\alpha)\right\}\\
\dot{\alpha}_k^* & = - i\frac{\partial\cH}{\partial\alpha_k} 
(\alpha^*,\alpha +p^*)\\
\dot p_k^* &= i \left\{ \frac{\partial\cH}{\partial\alpha_k^*} 
(\alpha^*,\alpha +p^*) - \frac{\partial\cH}{\partial\alpha_k^*} 
(\alpha^* +p,\alpha)\right\} 
\end{split} \right.
\end{equation}

We now fix $\alpha_0 \in \cee^N$ and solve the equation \eqref{eq:2.7}. 
Let 
\begin{equation*}
\begin{split}
\L (\alpha^*, \alpha, p^*,p) 
&= i\left\{ \left( 
p \frac{\partial\cH}{\partial \alpha^*} (\alpha^*+p,\alpha) 
- p^* \frac{\partial\cH}{\partial \alpha} (\alpha^*,\alpha +p^*)\right) 
\right.\\
&\qquad \left. \vphantom{\frac{\partial}{\partial}}
- \cH (\alpha^*+p,\alpha) + \cH(\alpha^*,\alpha +p^*) \right\}
\end{split}
\end{equation*}
Denote by $g^t$ the Hamiltonian dynamics corresponding to 
\begin{equation*}
\dot\alpha_k = i \frac{\partial\cH}{\partial\alpha_k^*} 
(\alpha^*,\alpha)\ ,\qquad k=1,\ldots,N\ .
\end{equation*}
Under the condition $|\cH (\alpha^*,\alpha)| \ge c_1|\alpha|^M -c_2$, 
where $M>0$, $c_1>0$, $c_2>0$ this dynamics is defined for all 
$\alpha (0) \in \real^N$, $t\in\real$
\begin{equation*}
\alpha = g^t (\alpha (0))
\end{equation*}
and is a 1-parametric group of (real)-analytic diffeomorphisms 
$\cee^N\simto \cee^N$.\footnote{To ensure the well-posedness of the evolution
\eqref{eq:1.3} one needs some additional assumptions on $\cH$. 
The following condition suffices (see \cite{Hor}, \cite{Shubin}).
There exist $R>0$, $\rho>0$ and a set of constants $C_{s\ell}$, 
where $s,\ell\in \zed_+^N$, so that 
$$\left|\frac{\partial^{|s+\ell|}}{\partial\alpha^{*s}\partial\alpha^\ell}
\cH (\alpha^*,\alpha)\right| 
\le C_{s,\ell} |\alpha|^{-\rho |s+\ell|}  |\cH (\alpha^*,\alpha)|\ ,
\qquad |\alpha|\ge R\ .$$} 

Let $\alpha$ belong to a small enough neighborhood of $g^t\alpha_0$. 
We solve \eqref{eq:2.9} with initial data
\begin{equation}\label{eq:2.10} 
\begin{cases}
\alpha_k(0)\ ,& k=1,\ldots,N\\
p_k(0) = - (\alpha_k^* (0) - \alpha_{0k}^*)\ ,& k=1,\ldots,N
\end{cases}
\end{equation}
Here $\alpha (0)$ is chosen in a small neighborhood of $\alpha_0$ in 
$\cee^N$. 

Linearizing \eqref{eq:2.9} along a trajectory with $p=0$, we obtain 
\begin{equation}\label{eq:2.11} 
\begin{cases}
\ds\dot A_k = i\left\{\frac{\partial^2\cH}{\partial\alpha_k^*\partial \alpha_r}
(\alpha^*,\alpha) A_r + 
\frac{\partial^2\cH}{\partial\alpha_k^*\partial\alpha_r^*} 
(\alpha^*,\alpha) (A_r^* + B_r)\right\} &\\
\dot B_k = i\left\{ \frac{\partial^2\cH}{\partial\alpha_k\partial\alpha_r} 
(\alpha^*,\alpha) B_r^* 
- \frac{\partial^2 \cH}{\partial\alpha_k\partial\alpha_r^*} 
(\alpha^*,\alpha)B_r\right\}\ ,&
\end{cases}
\end{equation}
where $A_k = \delta \alpha_k$, $B_k = \delta p_k$, $k=1,\ldots,N$.

{From} \eqref{eq:2.10} the initial condition for \eqref{eq:2.11} 
satisfies 
\begin{equation}\label{eq:2.12} 
B_k (0) = - A_k^* (0)\ ,\qquad k=1,\ldots,N\ .
\end{equation} 
We want to show that the map that assigns $\alpha (t)$ to each $\alpha (0)$ 
in a small neighborhood of $\alpha_0$ where
$(\alpha (t),p(t))$ solves \eqref{eq:2.9} with initial condition 
\eqref{eq:2.10} is a local diffeomorphism (of class $C^\omega$). 
Once this is established, $S$ can be defined in a small neighborhood 
of $g^t\alpha_0$ as follows: 
\begin{equation}\label{eq:2.13} 
S(\alpha^* (t),\alpha (t),t) = - |\alpha (0)-\alpha_0|^2 
+ \int_0^t \L (\alpha^* (\tau), \alpha (\tau),p^*(\tau),p(\tau))\,d\tau
\end{equation} 

Also, $p^*(t) = \frac{\partial S}{\partial\alpha^*} (\alpha^*(t),
\alpha(t),t)$, 
$p(t) = \frac{\partial S}{\partial\alpha} (\alpha^*(t),\alpha (t),t)$. 
To verify the statement above we notice that \eqref{eq:2.11} yields 
\begin{equation*}
\begin{split}
\frac{d}{dt} (A_kB_k + A_k^* B_k^*) 
& = i \left\{ 
\frac{\partial^2\cH}{\partial\alpha_k^*\partial\alpha_r} A_r B_k 
+\frac{\partial^2\cH}{\partial\alpha_k^*\partial\alpha_r^*} 
(A_r^* + B_r) B_k 
+ \frac{\partial^2\cH}{\partial \alpha_k\partial\alpha_r} A_k B_r^* 
- \frac{\partial^2\cH}{\partial \alpha_k\partial\alpha_r^*} A_k B_r\right\}\\
&\quad -i\left\{ 
\frac{\partial^2\cH}{\partial \alpha_r^*\partial\alpha_k} A_r^* B_k^* 
+\frac{\partial^2\cH}{\partial \alpha_r\partial\alpha_k} (A_r+ B_r^*) B_k^* 
+ \frac{\partial^2\cH}{\partial \alpha_r^*\partial\alpha_k^*} A_k^* B_r 
- \frac{\partial^2\cH}{\partial \alpha_r\partial\alpha_k^*} A_k^* B_r^* 
\right\}\\
& = i  
\frac{\partial^2\cH}{\partial \alpha_k^*\partial\alpha_r^*} B_r B_k 
- i \frac{\partial^2\cH}{\partial \alpha_r\partial\alpha_k} B_r^* B_k^*\\
& = - \frac{d}{dt} B_k B_k^*
\end{split}
\end{equation*} 

Integrating and using \eqref{eq:2.12} we get 
\begin{equation*}
A_k(t) B_k(t) + A_k^* (t) B_k^* (t) + B_k(t) B_k^* (t) 
= - 2|B(0)|^2 + |B(0)|^2 = - |B(0)|^2\ .
\end{equation*}
this implies $A_k(t) B_k(t) + A_k^* (t) B_k^* (t) \le - |B(t)|^2$. 
By using the second equation \eqref{eq:2.11} and \eqref{eq:2.12} we get 
$C|A(0)| \ge |B(t)| \ge C^{-1} |A(0)|$ with some $C>0$ uniform over any 
finite time interval $[0,T]$. 
Therefore, $|A(t)| \ge C^{-1} |A(0)|$ and one would never encounter a 
caustic moving along the trajectory \eqref{eq:2.9} with $p=0$ 
(i.e., along $g^t \alpha_0$). 
This is enough to conclude. 
The solution \eqref{eq:2.13} to \eqref{eq:2.7}, \eqref{eq:2.4} is 
unique in the obvious sense. 

Clearly $S((g^t \alpha_0)^*, g^t \alpha_0,t) =0$ and 
\begin{equation*}
\frac{\partial S}{\partial\alpha^*} ((g^t\alpha_0)^*, g^t\alpha_0,t)=0\ ,
\quad 
\frac{\partial S}{\partial\alpha} ((g^t\alpha_0)^*, g^t\alpha_0,t)=0\ .
\end{equation*}
Also, the function $S(\alpha^*,\alpha,t)$ satisfies the inequality 
\begin{equation*}
S(\alpha^*,\alpha,t) \le - C|\alpha - g^t \alpha_0|^2
\end{equation*}
in some possibly smaller neighborhood of $g^t \alpha_0$. 
This follows from the above argument since $A_k(t)B_k(t) + A_k^*(t) B_k^*(t)$ 
is the second differential of $S(t)$ at the critical point 
$((g^t\alpha_0)^*,g^t\alpha_0)$. 

\addtocounter{section}{1}
\setcounter{equation}{0}
\subsection*{3.}	
We now turn to the system of transport equations for the coefficients 
$b_j(\alpha^*,\alpha,t)$, $j\ge 0$. 
We have, substituting \eqref{eq:2.3} into \eqref{eq:1.10} and collecting 
terms of order $\hbar^0$: 
\begin{equation}\label{eq:3.1} 
\begin{split}
\dot b_0 & = i \sum_{r\in\zed_+^N} \sum_{\ell=1}^N \frac{r_\ell}{r!} 
\left\{ 
\Big(\frac{\partial}{\partial\alpha}\Big)^r
\cH(\alpha^*,\alpha) 
\left[
\Big(\frac{\partial S}{\partial\alpha^*}\Big)^{r-1_\ell} 
\frac{\partial}{\partial\alpha_\ell^*} 
b_0 + \frac12 
\frac{\partial}{\partial\alpha_\ell^*} 
\left( 
\Big(\frac{\partial S}{\partial\alpha^*}\Big)^{r-1_\ell} \right) b_0 \right]
\right.\\
&\qquad \left. 
- \Big(\frac{\partial}{\partial\alpha^*}\Big)^r 
\cH (\alpha^*,\alpha) 
\left[ 
\Big(\frac{\partial S}{\partial\alpha}\Big)^{r-1_\ell} 
\frac{\partial}{\partial\alpha_\ell} 
b_0 +\frac12 
\frac{\partial}{\partial\alpha_\ell} 
\left(\Big(\frac{\partial S}{\partial\alpha}\Big)^{r-1_\ell}\right) 
b_0 \right]\right\}\ .
\end{split}
\end{equation}
In other words, 
\begin{equation}\label{eq:3.2} 
\begin{split}
\dot b_0 & = i \sum_{\ell=1}^N \left\{ 
\frac{\partial}{\partial\alpha_\ell} 
\cH \Big( \alpha^*,\alpha + 
\frac{\partial S}{\partial\alpha^*} \Big)  
\frac{\partial}{\partial\alpha_\ell^*} 
- \frac{\partial}{\partial\alpha_\ell^*} 
\cH \Big( \alpha^* + 
\frac{\partial S}{\partial\alpha},\alpha \Big)  
\frac{\partial}{\partial\alpha_\ell}\right\} b_0 \\
&\qquad +\frac{i}2 \sum_{\ell=1}^N \sum_{m=1}^N \left\{ 
\frac{\partial^2}{\partial\alpha_\ell \partial\alpha_m} 
\cH \left(\alpha^*,\alpha + 
\frac{\partial S}{\partial\alpha^*} \right) 
\frac{\partial^2 S}{\partial\alpha_\ell^* \partial\alpha_m^*}
\right. \\
&\qquad \left.
-  \frac{\partial^2 }{\partial\alpha_\ell^* \partial\alpha_m^*} 
\cH \left(\alpha^* + \frac{\partial S}{\partial\alpha}, \alpha \right)
\frac{\partial^2 S}{\partial\alpha_\ell \partial\alpha_m} 
\right\} b_0\\
& \equiv L_0 b_0
\end{split}
\end{equation}
In the right side of \eqref{eq:3.1} $1_\ell$, $\ell =1,\ldots,N$ is 
the vector with components $\delta_{\ell s}$, $s=1,\ldots,N$. 

For an arbitrary $j\ge0$
\begin{equation}\label{eq:3.3} 
\dot b_j = L_0 b_j + \sum_{k=1}^j L_k b_{j-k}\ .
\end{equation} 
The differential operators $L_k$ in the right hand side of \eqref{eq:3.3} 
are as follows: 
\begin{equation}\label{eq:3.4} 
\begin{split}
L_k & = \frac{i}{(k+1)!} 
\frac{\partial^{k+1}}{\partial \hbar^{k+1}} 
\sum_{r\in\zed_+^N} \frac1{r!}
\left\{ 
\left( \frac{\partial}{\partial\alpha}\right)^r
\cH (\alpha^*,\alpha) 
\left(\hbar \frac{\partial}{\partial\alpha^*} + 
\frac{\partial S}{\partial\alpha^*} \right)^r 
\right.\\
&\qquad \left. -
\left(\frac{\partial}{\partial\alpha^*} \right)^r 
\cH (\alpha^*,\alpha) 
\left(\hbar \frac{\partial}{\partial\alpha} + 
\frac{\partial S}{\partial\alpha} \right)^r \right\}
\Big|_{\hbar=0}
\end{split}
\end{equation}
To solve the system \eqref{eq:3.1}--\eqref{eq:3.4} with initial conditions 
\eqref{eq:2.5}, \eqref{eq:2.6} we use \eqref{eq:2.9} to obtain with 
fixed $\alpha^* (0)$, $\alpha (0)$ in a small neighborhood of $\alpha_0 
\in \cee^N$ 
\begin{equation}\label{eq:3.5} 
\frac{d}{dt} b_0 (\alpha^* (t),\alpha^* (t),t) 
+ c_0 (\alpha^* (t),\alpha(t),t)b_0 =0
\end{equation}
where $c_0$ is the coefficient in the second term of \eqref{eq:3.2}. 
Having solved \eqref{eq:3.5} with initial condition \eqref{eq:2.5}, 
we obtain a first order linear inhomogeneous ODE for $b_1$, etc.. 
According to \eqref{eq:3.3} to solve the equation for $b_j$ we are 
recursively faced with the problem of solving the same ODE with a different 
known right hand side. 
The solution to this problem therefore exists and is unique for 
$(\alpha^*,\alpha)$ being sufficiently close to 
$((g^t\alpha_0)^*,g^t\alpha_0)$. 

To solve the problem \eqref{eq:1.9}, \eqref{eq:1.10} with the 
initial condition  
\begin{equation*} 
\langle \alpha |F(0)|\alpha\rangle = \alpha^{*m} \alpha^q
\end{equation*} 
we use the completeness relation 
\begin{equation*}
\frac1{(\pi\hbar)^N} \int |\alpha_0\rangle\ \langle\alpha_0| 
\, d^{2N} \alpha_0 = id
\end{equation*}
which leads to the expansion 
\begin{equation*}
\langle \alpha |F(t)|\alpha\rangle 
= \frac1{(\pi\hbar)^N} \int d^{2N} \alpha_0 
e^{\frac{S(\alpha^*,\alpha,t)}{\hbar}} 
\sum_{j=0}^\infty b_j (\alpha^*,\alpha,t) \hbar^j
\end{equation*} 
Here in the right hand side the phase $S$ and the coefficients $b_j$, 
$j\ge0$ depend on a parameter $\alpha_0 \in \cee^N$.

\addtocounter{section}{1}
\setcounter{equation}{0}
\subsection*{4. Example (see \cite{BIZ}, 
\cite{BBH} for the exact solution)} 

Let $N=1$, $\cH (\alpha^*,\alpha)=\omega \alpha^*\alpha +\mu\alpha^{*2}
\alpha^2$. 
The equation \eqref{eq:2.7} takes the form 
\begin{equation}\label{eq:4.1} 
\dot S = i(\omega + 2\mu |\alpha|^2)  
\left( \alpha^* \frac{\partial S}{\partial\alpha^*}
- \alpha \frac{\partial S}{\partial\alpha}\right) 
+ i\mu 
\left( \alpha^{*2} \left(\frac{\partial S}{\partial\alpha^*}\right)^2 
- \alpha^2 \left(\frac{\partial S}{\partial\alpha}\right)^2 \right)
\end{equation} 

The first transport equation is 
\begin{equation}\label{eq:4.2} 
\begin{split}
\dot b_0 
&= i(\omega +2\mu |\alpha|^2) 
\left( \alpha^* \frac{\partial}{\partial\alpha^*} 
- \alpha \frac{\partial}{\partial\alpha}\right) b_0 
+ 2i\mu \left( \alpha^{*2}
\frac{\partial S}{\partial \alpha^*}
\frac{\partial}{\partial \alpha^*} - 
\alpha^2 \frac{\partial S}{\partial \alpha}
\frac{\partial}{\partial \alpha}\right)  b_0\\ 
&\qquad + i\mu 
\left( \alpha^{*2} 
\frac{\partial^2 S}{\partial \alpha^{*2}} - \alpha^2 
\frac{\partial^2 S}{\partial \alpha^2}\right) b_0 
\equiv L_0 b_0
\end{split}
\end{equation}

A simple computation yields: 
\begin{align}
S(\alpha^*,\alpha,t) 
& = - |\alpha (0) - \alpha_0|^2 - i\mu t \left( \alpha^* (0)^2 \alpha_0^2 
- \alpha (0)^2 \alpha_0^{*2}\right)\label{eq:4.3}\\
&\qquad + |\alpha (0)|^2 \left( 1- e^{-2i\mu t(\alpha^* (0)\alpha_0 
- \alpha (0) \alpha_0^*)}\right)\notag\\
b_0 (\alpha^*,\alpha,t)
& = \alpha (0)^{*m} \alpha (0)^q 
\frac1{\sqrt{1+4\mu^2 t^2 |\alpha_0|^2 |\alpha (0)|^2}} \label{eq:4.4}\\
&\qquad \exp \left\{ 
\frac{i(\alpha (0)^*\alpha_0 - \alpha (0)\alpha_0^*)}
{2|\alpha (0)|\,|\alpha_0|} 
\arctan (2\mu t |\alpha_0|\, |\alpha (0)|) \right\}\ ,\notag
\end{align}
where 
\begin{equation}\label{eq:4.5} 
\alpha = \alpha (0) \exp \{ i\omega t + 2\mu t \alpha_0^*\alpha (0)\}\ .
\end{equation}
Notice that for any $t$ the map \eqref{eq:4.5} $\alpha (0)\to \alpha$ 
is invertible in a sufficiently small neighborhood of $\alpha_0$. 

\subsection*{Conclusion}

In conclusion, we developed a theory for Bose systems with many degrees 
of freedom which describes not an oscillating asymptotic behavior of 
a quasi-classical wave function, but a Laplace asymptotics for 
quantum average quantities, which are localized in phase space. 
This approach has many advantages in comparison with the standard one, 
as it does not require (i) an intermideate step of computing 
a complicated behaviour of a quasi-classical wave function which 
usually presents difficulties associated with caustics, 
and (ii) an additinal integration for computing the average values. 
Our approach
can be used for solving many dynamical  problems related to Bose systems 
with many degrees of freedom.

\subsection*{Acknowledgements}

This work was supported by the Department of Energy under the contract 
W-7405-ENG-36 and DOE Office of Basic Energy Sciences. 
The work of MV was supported in part by the National Science Foundation 
Grant DMS-9876947.
The work of GPB was partly supported by the National Security Agency (NSA) 
and by the Advanced Research and Development Activity (ARDA). 
MV is grateful to the Los Alamos National Laboratory 
for the  kind hospitality during his visit in July 2002. 

%

\end{document}